\documentclass[10pt,conference]{IEEEtran}

\usepackage{cite}
\usepackage{amsmath,amssymb,amsfonts}
\usepackage{graphicx}
\usepackage{textcomp}
\usepackage{xcolor}
\usepackage{booktabs}
\usepackage{multirow}
\usepackage{url}
\usepackage{hyperref}
\usepackage{balance}

\def\BibTeX{{\rm B\kern-.05em{\sc i\kern-.025em b}\kern-.08em
    T\kern-.1667em\lower.7ex\hbox{E}\kern-.125emX}}

\begin{document}

\title{Reducing Instruction-Fetch Energy in RISC-V for Embedded AI Processing via Dynamic and Static Loop Caching}

\author{
\IEEEauthorblockN{Wiebren Wijnstra, Sameed Sohail, Berend-Jan van der Zwaag, Sabih Gerez, Amirreza Yousefzadeh}
\IEEEauthorblockA{University of Twente, Enschede, The Netherlands}
}
\maketitle 

\begin{abstract}
Embedded RISC-V processors are increasingly deployed for on-device AI inference at the edge, where energy efficiency is a primary design constraint. Instruction fetching from SRAM-based memory is a dominant source of energy consumption in these cores, accounting for over 40\% of total energy in our baseline measurements. This paper presents two loop cache architectures integrated into the datapath of a RISC-V processor: a dynamic loop cache that automatically detects and caches short backward-branch loops at runtime, and a static loop cache that functions as a software-managed hot-code instruction buffer, allowing preloading of arbitrary instruction blocks during the boot sequence. Both designs are implemented in the open-source NEORV32 RISC-V processor and evaluated on a LeNet-5 convolutional neural network inference workload, synthesized on GlobalFoundries 22nm FDX+ technology at 0.5\,V and 250\,MHz. The dynamic cache reduces instruction fetches by 48.3\% and total energy by 21.5\%, while the static cache achieves an 83.3\% fetch reduction and 35.5\% total energy savings. The area overhead of both designs remains below 0.2\% of the full SoC area. The complete implementation is open source: \url{https://github.com/wwiebren/NEORV32_loopcache_for_SparkRV}.
\end{abstract}


\section{Introduction}

On-device AI inference is rapidly expanding into domains where energy budgets are measured in milliwatts: wearable health monitors, autonomous sensor nodes, and always-on keyword spotters. In these systems, the computation itself is only part of the energy story. Fetching the instructions that orchestrate that computation can dominate overall power consumption. Horowitz~\cite{horowitz2014energy} showed that the energy cost of accessing an on-chip SRAM is orders of magnitude higher than the arithmetic operation it feeds, and this disparity grows with memory size: a larger SRAM macro is more area-efficient per bit, but every read activates longer word lines and more bit-line capacitance, increasing the energy and latency of each individual access~\cite{yousefzadeh2025memory}. The classical remedy is a memory hierarchy that places small, fast storage close to the processor and relegates bulk storage to larger, slower arrays. In high-performance cores this hierarchy takes the form of multi-level caches with tag arrays, associativity logic, and replacement policies. For the small embedded processors that serve as controllers inside multi-core AI accelerators~\cite{abdallah2022neuromorphic, tang2023seneca, jianwei2023SFANC}, however, the silicon and energy overhead of a conventional cache can rival or exceed the overhead of the memory it is meant to protect~\cite{banakar2002scratchpad}.

The scale of the problem is significant. Measurements on embedded RISC-V cores running general workloads consistently attribute 30\% to 50\% of total chip power to instruction-memory access~\cite{lee1999low, segars2001low, gordon2003tiny, tang2023seneca}. Our own baseline measurements on an AI inference workload (Section~\ref{sec:results}) confirm this: instruction memory accounts for 40.2\% of total energy, making it the single largest contributor, ahead of the ALU and data memory combined. Any mechanism that can serve instructions from a smaller, lower-capacitance storage without the overhead of a full cache therefore translates directly into whole-system energy savings.

Loop caches offer exactly this opportunity. Because embedded software, and neural-network inference code in particular, spends the vast majority of its execution time inside loops, a small register-based buffer that captures and replays loop bodies can intercept a large fraction of instruction fetches at a fraction of the energy cost of an SRAM read. Unlike a conventional cache, a loop cache requires no tag storage, no associativity logic, and no replacement policy; it simply holds a contiguous block of instructions and checks whether the current program counter falls within the cached range. Two flavors have been proposed in the literature. \textit{Dynamic} loop caches, introduced by Lee~et~al.~\cite{lee1999instruction, lee1999low}, monitor the instruction stream at runtime for short backward branches (SBBs), fill a small buffer during the second iteration of a detected loop, and serve subsequent iterations entirely from that buffer, suppressing SRAM accesses. The approach is fully transparent to software and requires no compiler or ISA support, but it is restricted to tight, well-formed loops that contain no internal branches, nested loops, or subroutine calls~\cite{dipasquale2003hardware}. \textit{Static} loop caches, proposed by Gordon-Ross~et~al.~\cite{gordon2003tiny}, take a semi-software-controlled approach: during the boot sequence, application code preloads the buffer with one or more contiguous instruction regions identified through prior profiling and configures Loop Address Registers (LARs) with their start and end addresses. At runtime, a set of parallel comparators continuously checks whether the program counter falls within any registered region. Because the cached unit is a contiguous code block rather than individual cache lines, a static loop cache can capture large, complex code regions, including those with nested loops, internal branches, and subroutine calls, that would defeat a dynamic detector. The trade-off is the up-front profiling step needed to select the most frequently executed regions.

Both approaches differ fundamentally from conventional instruction caches such as the filter cache of Kin~et~al.~\cite{kin1997filter}, which still relies on tag comparisons and set-associative lookup on every access. They also differ from hardware loop support in platforms such as PULP~\cite{gautschi2017near}, where dedicated loop-count registers and zero-overhead loop instructions are added to the ISA, requiring a custom compiler toolchain. The PULP cores additionally use a small L0 instruction buffer in the fetch stage to reduce pressure on the shared instruction cache, but this buffer does not suppress SRAM accesses for arbitrary hot-code regions the way a static loop cache does. Compared with a full instruction scratchpad~\cite{banakar2002scratchpad}, the static loop cache presented here is far smaller (tens of instructions rather than kilobytes) and is accessed only through PC-range matching for selected hot regions, avoiding the need for explicit address remapping by the compiler.

Despite the potential of loop caching, the existing evaluations~\cite{lee1999instruction, gordon2003tiny} were conducted on general embedded benchmarks whose loop structures differ substantially from those found in AI inference workloads. Neural-network kernels, especially convolution, pooling, and dense-layer routines implemented in software on microcontrollers, are characterized by deeply nested loops with large iteration counts and frequent calls to arithmetic subroutines. These characteristics challenge the tight-loop assumption of dynamic caches and, at the same time, create an ideal target for a static cache that can cover an entire hot-code region in a single buffer entry. A systematic evaluation of both loop-caching strategies on an embedded AI workload, integrated directly into the processor datapath, has not been reported.

This paper addresses that gap. We present the design, implementation, and evaluation of both a dynamic and a static loop cache integrated into the datapath of the open-source NEORV32 RISC-V processor~\cite{NEORV32_RISCV}. Our contributions are:

\begin{itemize}
    \item We integrate two low-overhead instruction-reuse mechanisms into the NEORV32 fetch/decode path: a dynamic SBB-based loop cache and a software-managed static hot-code buffer with PC-range matching. Both implementations are fully open source\footnote{\url{https://github.com/wwiebren/NEORV32_loopcache_for_SparkRV}}. The main instruction memory is implemented with SRAM (larger, more area-efficient), while the loop cache storage uses registers (smaller, more energy-efficient per access).

    \item We show how placing the loop cache between the Instruction Prefetch Buffer (IPB) and the Execute Engine enables transparent instruction delivery and fetch suppression without any modification to the RISC-V ISA or compiler toolchain.

    \item We evaluate both designs on a LeNet-5 convolutional neural network inference workload synthesized in GlobalFoundries 22nm FDX+ at 0.5\,V and 250\,MHz, demonstrating total energy reductions of 21.5\% (dynamic) and 35.5\% (static) with an area overhead below 0.2\% of the full SoC.

    \item We conduct a parameter sweep and analyze why the static mechanism is better suited to the evaluated workload: it captures hot regions containing nested loops, internal branches, and subroutine calls, whereas the dynamic mechanism is restricted to well-formed tight loops.
\end{itemize}

\section{Proposed Architecture}\label{sec:architecture}

Both loop cache designs are integrated into the NEORV32 RISC-V processor~\cite{NEORV32_RISCV}, an open-source, customizable System-on-Chip built around a two-stage pipelined multi-cycle CPU implementing the RV32I base instruction set with optional extensions. The key architectural idea shared by both designs is the same: a small register-based instruction buffer is inserted into the processor pipeline at a point where it can transparently intercept instruction delivery and, when it holds the instructions the processor needs, suppress all accesses to the main SRAM-based instruction memory.

\subsection{Pipeline Integration}

Both loop cache variants sit between the Instruction Prefetch Buffer (IPB) and the Execute Engine, as shown in Fig.~\ref{fig:dynamic_arch}. This location is a natural fit for three reasons. First, the Execute Engine already exposes all the signals needed for loop detection: the current program counter, the decoded opcode, branch immediates, and the branch-taken flag. Second, the valid/acknowledge/data handshake between the IPB and the Execute Engine can be intercepted cleanly, allowing the loop cache to substitute its own instruction data without any changes to either the upstream Fetch Engine or the downstream execution logic. Third, when the loop cache is serving instructions, it asserts a \texttt{fetch\_disable} signal that halts the Fetch Engine and suppresses instruction-memory read accesses, directly eliminating fetch energy.

When the loop cache is inactive, it passes all bus signals through transparently, making it invisible to the rest of the pipeline. When active, it drives the data line from its internal register file and suppresses the IPB valid signal, stalling the Fetch Engine. In the current implementation, fetch suppression eliminates dynamic IMEM read energy during loop-cache hits. A \texttt{decode\_valid} register is added to the Execute Engine so that the loop cache evaluates CPU signals only once per decoded instruction, accounting for the multi-cycle nature of the NEORV32 pipeline.

\subsection{Dynamic Loop Cache}

\begin{figure*}
\centering
\includegraphics[width=\textwidth]{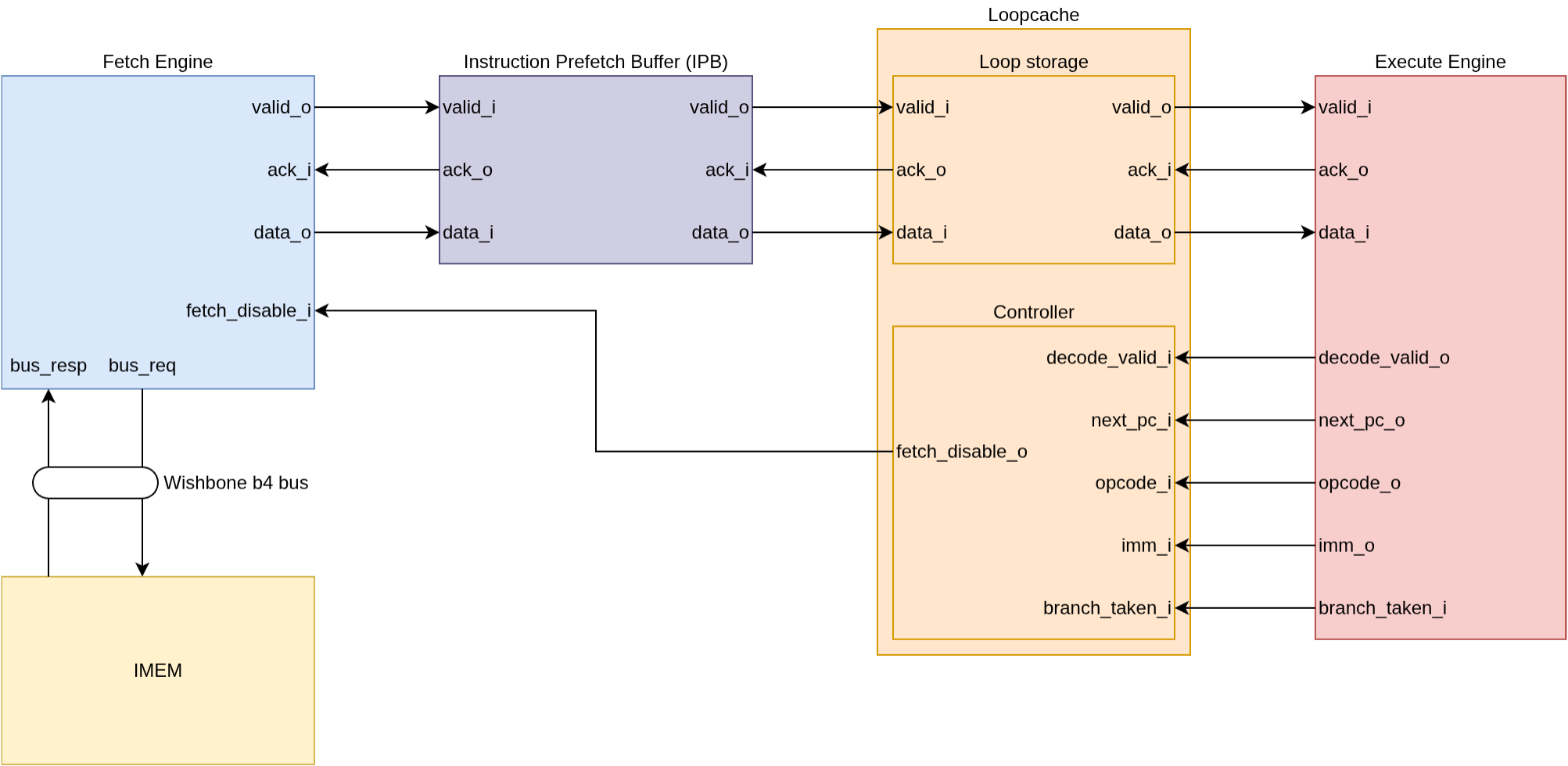}
\caption{Dynamic loop cache architecture integrated into the NEORV32 RISC-V processor pipeline. The Loopcache module (orange) sits between the Instruction Prefetch Buffer (IPB, purple) and the Execute Engine (red). A Loop Storage unit intercepts the instruction data path (valid/ack/data handshake), while a Controller monitors decode-stage signals from the Execute Engine. When active, the Controller asserts \texttt{fetch\_disable\_o} to halt the Fetch Engine (blue) and suppress instruction memory (Yellow, IMEM) read accesses.}
\label{fig:dynamic_arch}
\end{figure*}

The dynamic loop cache detects and captures short loops entirely in hardware, requiring no software support. Its design follows the Short Backward Branch (SBB) detection scheme of Lee~et~al.~\cite{lee1999low}. As shown in Fig.~\ref{fig:dynamic_arch}, the Loopcache module consists of a Loop Storage unit and a Controller, inserted between the IPB and Execute Engine. The Controller monitors the instruction stream through five signals from the Execute Engine: \texttt{decode\_valid}, \texttt{next\_pc}, \texttt{opcode}, \texttt{imm}, and \texttt{branch\_taken}.

Detection and filling are managed by a three-state finite state machine. In the \textsc{idle} state, the controller watches for taken backward branches whose displacement fits within the cache size, ensuring the entire loop body can be captured. When such an SBB is detected, the controller transitions to \textsc{fill} and records every instruction fetched during the next loop iteration into a direct-mapped register file, starting from the backward-branch target address. If the same SBB is taken again without any intervening change-of-flow (CoF) instruction, the fill is complete and the controller enters the \textsc{active} state, redirecting all subsequent instruction fetches to the register file. The cache remains active until a CoF instruction other than the triggering SBB is encountered, at which point the controller returns to \textsc{idle}. Because the cache is indexed by the lower bits of the instruction address, address translation is zero-overhead.

The strength of this design is its simplicity: it is a drop-in hardware module that captures the innermost tight loops automatically. Its limitation is equally clear: any loop body that contains an internal branch, a nested loop, or a subroutine call will trigger a CoF event during the fill phase and abort the caching attempt.

\subsection{Static Loop Cache}

\begin{figure*}[t]
\centering
\includegraphics[width=\textwidth]{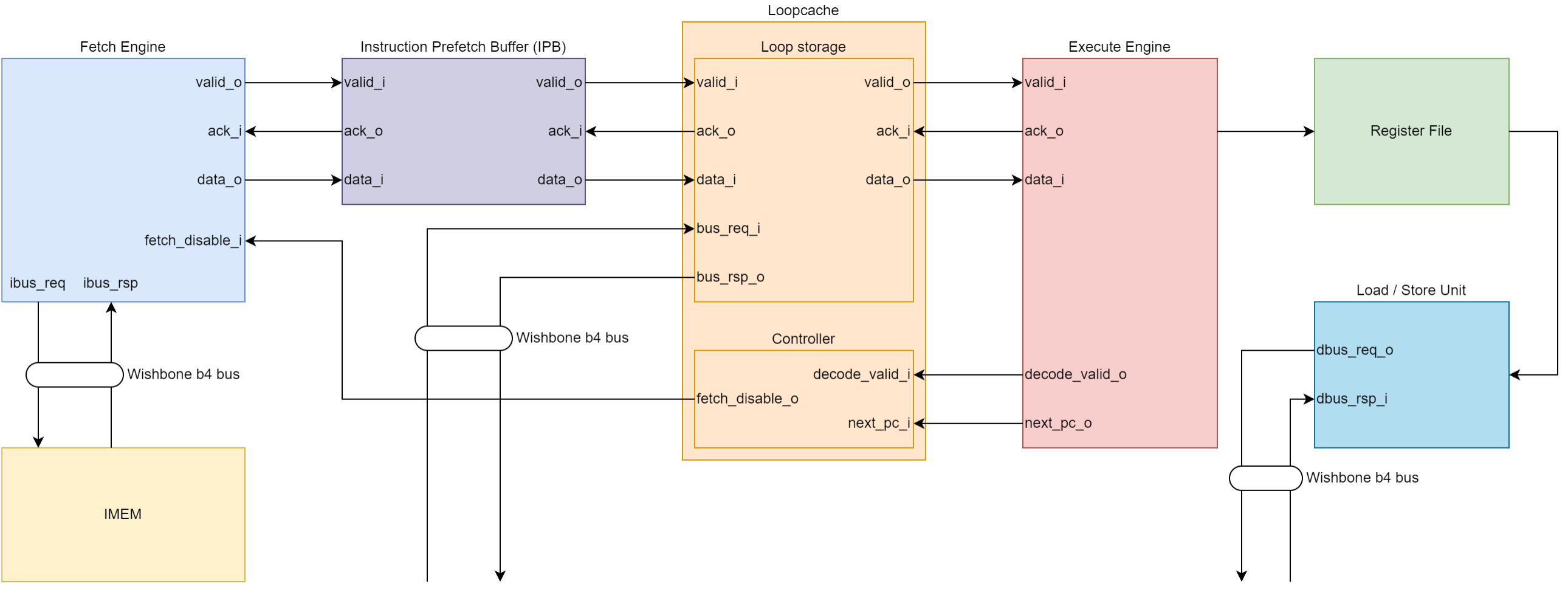}
\caption{Static loop cache architecture. Compared to the dynamic variant (Fig.~\ref{fig:dynamic_arch}), the Loop Storage gains a Wishbone b4 bus interface (bus\_req\_i/bus\_rsp\_o) for memory-mapped writes from the CPU's Load/Store Unit. The Controller requires fewer status signals (only decode\_valid and next\_pc) since loop boundaries are configured by software rather than detected at runtime. The Register File and Load/Store Unit within the Execute Engine are shown explicitly.}
\label{fig:static_arch}
\end{figure*}

The static loop cache removes the structural restrictions of the dynamic variant by shifting the responsibility for selecting what to cache from runtime hardware detection to a one-time software configuration step. Fig.~\ref{fig:static_arch} shows the modified architecture. The Loop Storage is now additionally connected to the Wishbone b4 data bus through bus\_req/bus\_rsp ports, making it writable from the CPU. During the boot sequence, application code copies the desired instruction regions from main memory into the loop storage and programs a set of Loop Address Registers (LARs) with the corresponding address ranges.

\begin{figure}[!b]
    \centering
    \includegraphics[width=0.9\columnwidth]{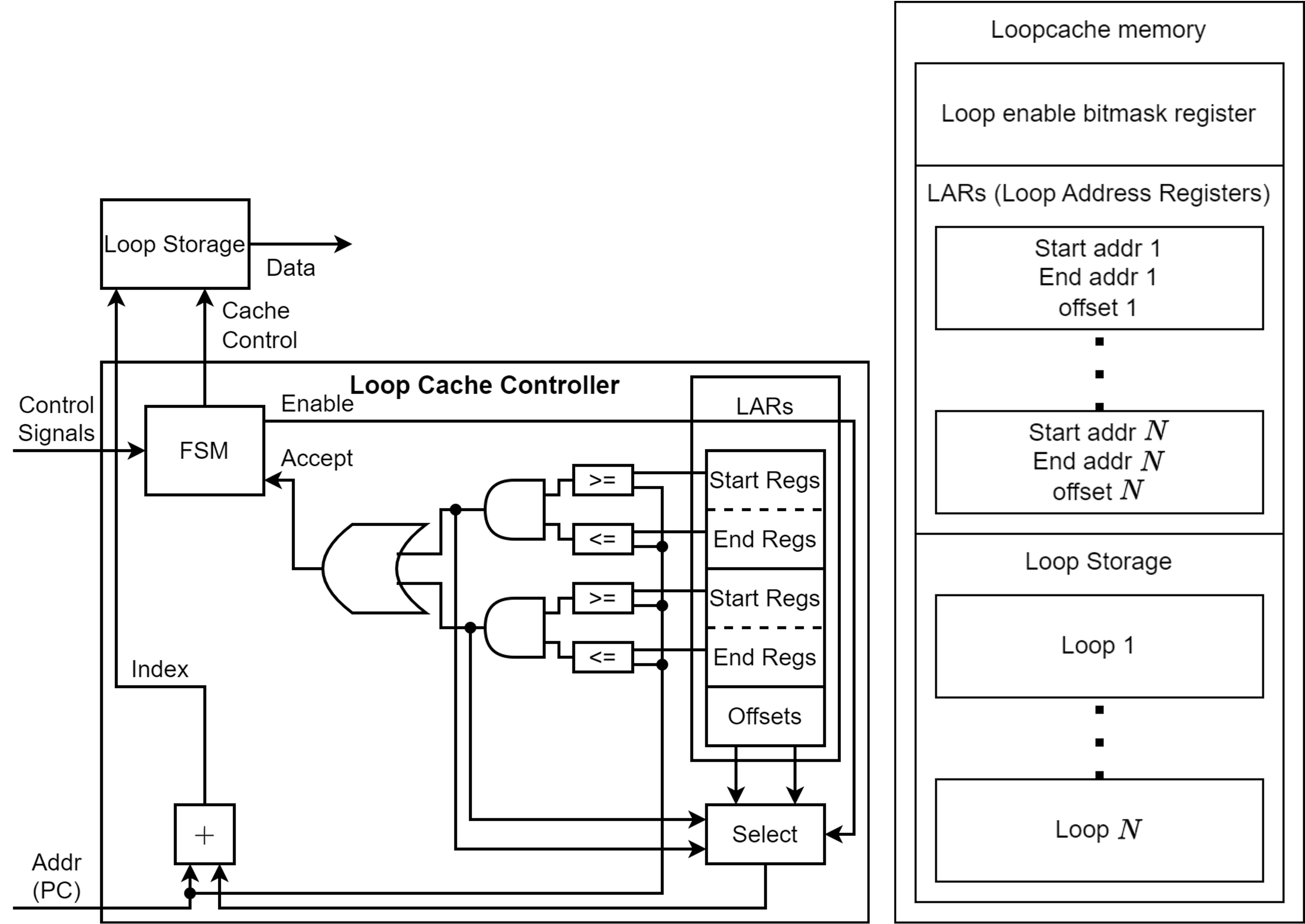}
    \caption{Internal architecture of the static loop cache controller. The FSM manages cache state while the LARs store start/end address pairs and precomputed offsets for each cached region. Parallel comparators check whether the current PC falls within any cached region, and the selected offset is added to compute the loop storage index.}
    \label{fig:controller}
\end{figure}

The cache consists of three components: a loop storage register file, the LARs, and a fully combinational controller whose internal architecture is shown in Fig.~\ref{fig:controller}. Each LAR holds a start address, an end address, and a precomputed offset defined as $\text{offset} = -M + C$, where $M$ is the region's start address in main memory and $C$ is its start address in the loop storage. At runtime, the controller compares the current program counter against all enabled LARs in parallel using greater-than-or-equal and less-than-or-equal comparators for the start and end boundaries, respectively. When the PC falls within a registered region, the corresponding offset is selected and added to the PC to produce the loop storage index: $\text{LPC} = \text{PC} + \text{offset}$. An enable bitmask allows individual regions to be activated or deactivated, saving comparator switching energy for unused entries.

Although we refer to this structure as a static loop cache for consistency with prior work~\cite{gordon2003tiny}, it is more accurately described as a software-managed hot-code instruction buffer. Any contiguous instruction region can be mapped into the local storage and selected through PC-range matching, regardless of its internal control flow. A single LAR may cover a code region that spans multiple nested loops and subroutine calls; therefore, the number of LARs and the number of syntactic loops in the cached code are not necessarily equal. This flexibility is precisely what makes the static cache well suited to AI inference workloads, where the hot code typically consists of a small number of deeply nested kernel routines (convolution, pooling, dense-layer multiply-accumulate) that together account for the overwhelming majority of executed instructions.

The one requirement is that the hot regions must be identified before deployment. For AI workloads, this is rarely a burden in practice: the computational kernels are known at compile time, their loop structure is regular and well-understood, and a single profiling run on representative input is sufficient to rank code regions by execution frequency. To support this step, we implement a lightweight hardware loop profiler that counts backward-branch executions during simulation. The profiler is not part of the final deployed hardware; it is used only during development to collect branch-frequency and PC-range data. Candidate hot-code regions are then selected offline and loaded into the static cache at boot time through the memory-mapped interface. The content of the static cache can be updated at runtime (via RISC-V), which is useful when the cache size is insufficient to cover all hot loops. 

\section{Experimental Results}\label{sec:results}

\subsection{Experimental Setup}

All experiments compare the two loop cache architectures against a baseline NEORV32 configuration with no loop cache. The entire design flow targets GlobalFoundries 22nm FDX+ technology using the low-power standard-cell library at 0.5\,V and 25$^\circ$C, operating at 250\,MHz. Instruction and data memories are standard-cell SRAMs generated with Synopsys Memory Compiler~\cite{synopsys_memory_compiler}, organized as 8192$\times$32-bit cells with bank-level sleep-mode support. Logic synthesis and area reporting use Cadence Genus~\cite{cadence_genus}, cycle-accurate waveforms are produced with Cadence Xcelium~\cite{cadence_xcelium}, and post-synthesis power estimation is performed with Cadence Joules~\cite{cadence_joules}. Because these estimates are pre-layout, they do not capture routing parasitics or clock-tree effects; the absolute energy figures should therefore be read as comparative estimates rather than silicon-accurate measurements.

The evaluation benchmark is a modified LeNet-5~\cite{lenet5} convolutional neural network whose configuration is summarized in Table~\ref{tab:benchmark}. The network contains two convolutional layers, two max-pooling layers, and three fully connected layers, totaling 61,706 parameters. Inference is executed in single-precision floating point via the Zfinx extension with the hardware multiplier enabled. Weights reside in a 1\,MB DMEM, and instructions occupy a 32\,KB IMEM. The inference code is compiled from C, and the same code runs on all three configurations; the only difference is a short boot-time initialization routine that preloads the static cache. Each benchmark run performs a single inference bracketed by two 1\,ms sleep cycles to capture a realistic power profile. Real edge deployments typically use integer-quantized inference, but floating-point execution exercises the instruction-fetch path more heavily due to library-level arithmetic routines, making this a conservative test of the loop cache's effectiveness.

\begin{table}[t]
\centering
\caption{LeNet-5 benchmark configuration.}
\label{tab:benchmark}
\footnotesize
\setlength{\tabcolsep}{4pt}
\begin{tabular}{@{}ll@{}}
\toprule
\textbf{Parameter} & \textbf{Value} \\
\midrule
Model & Modified LeNet-5 (61,706 parameters) \\
Input & 28$\times$28 grayscale \\
Precision & Single-precision float (Zfinx) \\
ISA & RV32IMF\_Zfinx \\
IMEM / DMEM & 32\,KB / 1\,MB \\
Same code across designs & Yes (except static cache init) \\
\bottomrule
\end{tabular}
\end{table}

The default size of both loop caches is 32 instructions (32$\times$32-bit). The static cache is populated with the most frequently executed code regions as identified by the simulation-time hardware profiler described in Section~\ref{sec:architecture}. All reported energy figures cover inference only; the one-time boot-time preload overhead is excluded, as it is amortized across the many inferences a deployed edge device performs over its lifetime.

\subsection{Baseline Energy Breakdown}

\begin{figure}[t]
    \centering
    \includegraphics[width=0.4\textwidth]{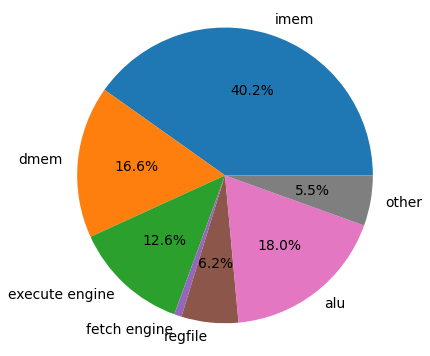}
    \caption{Baseline energy breakdown for LeNet-5 inference on the NEORV32 SoC.}
    \label{fig:baseline_energy}
\end{figure}

Fig.~\ref{fig:baseline_energy} breaks down the energy consumed by a single LeNet-5 inference on the baseline processor. Instruction memory dominates at 40.2\% of the total, followed by the ALU at 18\% and DMEM at 16.6\%. Even with the hardware multiplier and floating-point unit active, fetching instructions costs more than executing them. This confirms the motivation laid out in the introduction and establishes instruction-fetch suppression as the highest-leverage optimization target.

\subsection{Energy Reduction with Loop Caching}

\begin{table}[t]
\centering
\caption{Loop cache results for LeNet-5 (32-instruction storage).}
\label{tab:results}
\footnotesize
\setlength{\tabcolsep}{4pt}
\begin{tabular}{@{}lccc@{}}
\toprule
\textbf{Metric} & \textbf{Baseline} & \textbf{Dynamic} & \textbf{Static} \\
\midrule
Total energy ($\mu$J) & 116.89 & 91.81 ($\downarrow$21.5\%) & 75.39 ($\downarrow$35.5\%) \\
\midrule
IMEM energy ($\mu$J) & 46.96 & 20.83 ($\downarrow$55.6\%) & 6.64 ($\downarrow$85.9\%) \\
\midrule
Instr.\ fetches (count) & 7,177,817 & 3,711,891 ($\downarrow$48.3\%) & 1,202,303 ($\downarrow$83.3\%) \\
\midrule
Inference time (ms) & 132.94 & 130.61 ($\downarrow$1.75\%) & 128.36 ($\downarrow$3.44\%) \\
\bottomrule
\end{tabular}
\end{table}

\begin{figure}[t]
    \centering
    \includegraphics[width=\columnwidth]{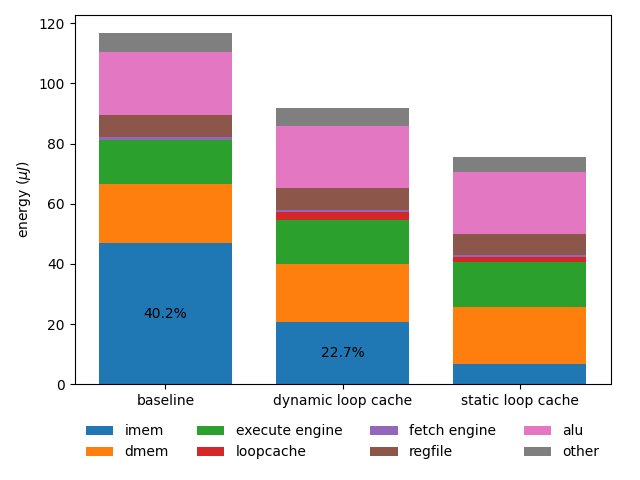}
    \caption{Per-component energy breakdown across all three configurations. The IMEM contribution (blue) shrinks substantially with both loop cache variants.}
    \label{fig:energy_comparison}
\end{figure}

Table~\ref{tab:results} and Fig.~\ref{fig:energy_comparison} present the results for both loop caches at the same 32-instruction storage size. The dynamic cache eliminates 48.3\% of all instruction fetches and reduces total inference energy by 21.5\%. The static cache goes considerably further, suppressing 83.3\% of fetches and saving 35.5\% of total energy. Looking at instruction-memory energy in isolation, the reductions are even more striking: 55.6\% for the dynamic cache and 85.9\% for the static cache.

The performance gap between the two designs is explained by the nature of the LeNet-5 code. The dynamic cache can only capture the innermost tight loops that contain no internal branches or subroutine calls. In a CNN inference workload, however, the hottest code regions are deeply nested kernel routines (convolution, pooling, dense-layer multiply-accumulate) that include function calls and conditional logic within the loop body. The static cache captures these entire regions as contiguous blocks, regardless of their internal control flow, and therefore intercepts a much larger share of the executed instruction stream.

Both caches also yield a modest reduction in inference time (up to 3.44\% for the static cache). This speedup is a side effect of eliminating fetch-side stalls and handshake overhead that occur when the same loop body is fetched repeatedly from SRAM.

\subsection{Area Overhead}

At the 32-instruction configuration, the dynamic loop cache occupies approximately 1,200\,$\mu\text{m}^2$ and the static loop cache approximately 2,650\,$\mu\text{m}^2$. The larger footprint of the static variant comes from the memory-mapped bus interface and the LAR comparison logic. Relative to the full NEORV32 SoC (including 1\,MB DMEM and 32\,KB IMEM), both figures fall below 0.2\% of total chip area. Because this SoC is heavily memory-dominated, the percentage relative to the CPU logic alone (excluding SRAM macros) is proportionally larger, but remains modest in absolute terms.

\subsection{Parameter Sweep}

\begin{figure}
    \centering
    \includegraphics[width=\columnwidth]{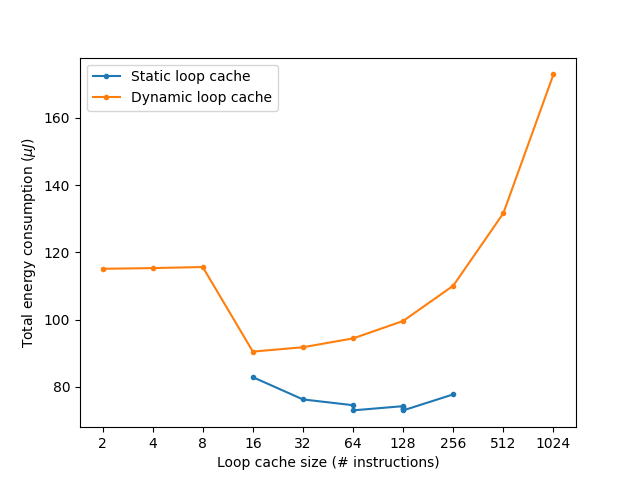}
    \caption{Total inference energy as a function of loop cache size for both architectures. The static cache continues to benefit from larger storage, while the dynamic cache saturates at 16 instructions and begins to cost more energy than it saves beyond 32.}
    \label{fig:size_vs_energy}
\end{figure}

To understand how each architecture scales, we sweep the cache size across a wide range on the LeNet-5 benchmark. Fig.~\ref{fig:size_vs_energy} shows the results.

\textbf{Dynamic cache.} Cache sizes from 2 to 1024 instructions are evaluated. A sharp improvement appears at 16 instructions, the point at which the most frequently executed innermost loop fits entirely in the buffer. Beyond 32 instructions no additional fetches are eliminated, confirming that the LeNet-5 code contains no cacheable well-formed loops larger than this. For sizes above 32, total energy actually \emph{increases} because the register file itself consumes more switching energy than the avoided SRAM accesses save.

\textbf{Static cache.} Starting from the single hottest code region (16 instructions), successively less-frequent regions are added. The first region alone eliminates 66.3\% of instruction fetches and reduces total energy by 29.1\%. Four code regions (55 instructions total, fitting in a 64-entry storage with 2 LARs, since a single LAR can span a contiguous region containing multiple syntactic loops) reach the sweet spot: 93.8\% fewer instruction fetches and 37.5\% energy savings. Beyond this point, each additional region yields diminishing returns that are increasingly offset by the energy cost of the larger register file.

Note that Table~\ref{tab:results} compares both caches at an equal 32-instruction size for a fair head-to-head comparison, while this sweep explores the optimal configuration for each architecture independently. The optimum cache size is application-dependent. However, as shown, the static cache consistently outperforms the dynamic cache across all sizes and continues to benefit from additional storage, whereas the dynamic cache saturates early and eventually becomes counterproductive.

\section{Conclusion}

This paper presented two loop cache architectures to reduce instruction fetch energy in a RISC-V-based embedded AI processor. Both designs are integrated directly into the processor datapath between the instruction prefetch buffer and the execute engine, enabling transparent instruction delivery and fetch suppression during cache hits.

Evaluation on a LeNet-5 inference workload synthesized in GF 22nm FDX+ at 0.5\,V demonstrates that the static loop cache is the more effective mechanism for this workload, achieving a 35.5\% total energy reduction and an 83.3\% instruction fetch elimination with an area overhead below 0.2\% of the full SoC. The dynamic loop cache, while limited to well-formed non-nested loops, offers a plug-and-play alternative that requires no software modifications or profiling and still achieves a 21.5\% energy reduction.

For system designers, the practical takeaway is that the choice between the two architectures depends on the workload and the development flow. When loop structures are simple and profiling is not feasible, the dynamic cache provides meaningful savings with zero software effort. When the workload is known in advance and hot-code regions can be identified through profiling, the static cache delivers substantially better results. The two approaches could also be combined in a hybrid architecture.

Both implementations are open source and available at \url{https://github.com/wwiebren/NEORV32_loopcache_for_SparkRV}. Future work includes developing automated profiling tools to simplify static cache configuration, evaluating the approach on additional AI workloads (including quantized models and transformer-based architectures), and validating the results through ASIC fabrication.

\section*{Acknowledgment}
We kindly acknowledge EUROPRACTICE for its support with the design tool.

\bibliographystyle{IEEEtran}
\bibliography{references}

@article{gordon2003tiny,
  title={Tiny instruction caches for low power embedded systems},
  author={Gordon-Ross, Ann and Cotterell, Susan and Vahid, Frank},
  journal={ACM Transactions on Embedded Computing Systems (TECS)},
  volume={2},
  number={4},
  year={2003},
  publisher={ACM New York, NY, USA}
}

@inproceedings{segars2001low,
  title={Low power design techniques for microprocessors},
  author={Segars, Simon},
  booktitle={International Solid-State Circuits Conference Tutorial},
  year={2001}
}

@article{lee1999low,
  title={Low-cost embedded program loop caching-revisited},
  author={Lee, Lea Hwang and Moyer, Bill and Arends, John and Arbor, Ann},
  journal={University of Michigan Technical Report CSE-TR-411-99},
  year={1999}
}

@inproceedings{lee1999instruction,
  title={Instruction fetch energy reduction using loop caches for embedded applications with small tight loops},
  author={Lee, Lea Hwang and Moyer, Bill and Arends, John},
  booktitle={Proceedings of the 1999 International Symposium on Low Power Electronics and Design},
  year={1999}
}

@book{abdallah2022neuromorphic,
  title={Neuromorphic Computing Principles and Organization},
  author={Abdallah, Abderazek Ben and Dang, Khanh N},
  year={2022},
  publisher={Springer}
}

@article{tang2023seneca,
  title={{SENECA}: building a fully digital neuromorphic processor, design trade-offs and challenges},
  author={Tang, Guangzhi and Vadivel, Kanishkan and Xu, Yingfu and Bilgic, Refik and Shidqi, Kevin and Detterer, Paul and Traferro, Stefano and Konijnenburg, Mario and Sifalakis, Manolis and van Schaik, Gert-Jan and others},
  journal={Frontiers in Neuroscience},
  volume={17},
  year={2023},
  publisher={Frontiers Media SA}
}

@misc{NEORV32_RISCV,
  author = {Nolting, Stephan and {All The Awesome Contributors}},
  doi = {10.5281/zenodo.13872735},
  title = {{The NEORV32 RISC-V Processor}},
  url = {https://github.com/stnolting/neorv32},
  year = {2023}
}

@article{jianwei2023SFANC,
  author={Jianwei, Xue and Rendong, Ying and Faquan, Chen and Peilin, Liu},
  journal={IEEE Transactions on Very Large Scale Integration (VLSI) Systems},
  title={{SFANC}: Scalable and Flexible Architecture for Neuromorphic Computing},
  year={2023},
  volume={31},
  number={11},
  doi={10.1109/TVLSI.2023.3282239}
}

@inproceedings{kin1997filter,
  author={Kin, J. and Gupta, Munish and Mangione-Smith, W.H.},
  booktitle={Proceedings of 30th Annual International Symposium on Microarchitecture},
  title={The filter cache: an energy efficient memory structure},
  year={1997},
  doi={10.1109/MICRO.1997.645809}
}

@article{lenet5,
  title={Gradient-based learning applied to document recognition},
  author={LeCun, Yann and Bottou, L{\'e}on and Bengio, Yoshua and Haffner, Patrick},
  journal={Proceedings of the IEEE},
  volume={86},
  number={11},
  year={1998},
  publisher={IEEE}
}

@article{gautschi2017near,
  author={Gautschi, Michael and Schiavone, Pasquale Davide and Traber, Andreas and Loi, Igor and Pullini, Antonio and Rossi, Davide and Flamand, Eric and G\"{u}rkaynak, Frank K. and Benini, Luca},
  journal={IEEE Transactions on Very Large Scale Integration (VLSI) Systems},
  title={Near-Threshold {RISC-V} Core With {DSP} Extensions for Scalable {IoT} Endpoint Devices},
  year={2017},
  volume={25},
  number={10},
  doi={10.1109/TVLSI.2017.2654506}
}

@misc{synopsys_memory_compiler,
  title = {{Synopsys Memory Compiler IP}},
  author={{Synopsys}},
  howpublished = {\url{https://www.synopsys.com}},
  year = {2025}
}

@misc{cadence_xcelium,
  title = {{Cadence Xcelium Simulator}},
  author = {{Cadence Design Systems, Inc.}},
  year = {2023}
}

@misc{cadence_joules,
  title = {{Cadence Joules RTL Power Solution}},
  author = {{Cadence Design Systems, Inc.}},
  year = {2015}
}

@misc{cadence_genus,
  title = {{Cadence Genus Synthesis Solution}},
  author = {{Cadence Design Systems, Inc.}},
  year = {2015}
}

@misc{dipasquale2003hardware,
  title={Hardware Loop Buffering},
  author={DiPasquale, Scott and Elmeleegy, Khaled and Ganier, CJ and Swanson, Erik},
  year={2003},
  publisher={Citeseer}
}

@inproceedings{horowitz2014energy,
  author={Horowitz, Mark},
  booktitle={2014 IEEE International Solid-State Circuits Conference Digest of Technical Papers (ISSCC)},
  title={1.1 Computing's energy problem (and what we can do about it)},
  year={2014},
  pages={10--14},
  doi={10.1109/ISSCC.2014.6757323},
  publisher={IEEE}
}

@inproceedings{banakar2002scratchpad,
  author={Banakar, Rajeshwari and Steinke, Stefan and Lee, Bo-Sik and Balakrishnan, M. and Marwedel, Peter},
  booktitle={Proceedings of the Tenth International Symposium on Hardware/Software Codesign (CODES)},
  title={Scratchpad memory: a design alternative for cache on-chip memory in embedded systems},
  year={2002},
  pages={73--78},
  doi={10.1145/774789.774805},
  publisher={ACM}
}

@inproceedings{yousefzadeh2025memory,
  title={Memory Wall is not gone: A Critical Outlook on Memory Architecture in Digital Neuromorphic Computing},
  author={Yousefzadeh, Amirreza and Sohail, Sameed and Varbanescu, Ana Lucia},
  booktitle={2025 IEEE Computer Society Annual Symposium on VLSI (ISVLSI)},
  volume={1},
  pages={1--4},
  year={2025},
  organization={IEEE}
}

\end{document}